# Neutrino-4 experiment on search for sterile neutrino with multi-section model of detector

A.P. Serebrov[a*], V.G. Ivochkin[a], R.M. Samoilov[a], A.K. Fomin[a], A.O. Polyushkin[a], V.G. Zinoviev[a], P.V. Neustroev[a], V.L. Golovtsov[a], A.V. Chernyj[a], O.M. Zherebtsov[a], V.P. Martemyanov[b], V.G. Tarasenkov[b], V.I. Aleshin[b], A.L. Petelin[c], A.L. Ishutov[c], A.A. Tuzov[c], S.A. Sazontov[c], D.K. Ryazanov[c], M.O. Gromov[c], V.V. Afanasiev[c], M.E. Zaytsev[a], M.E. Chaikovskii[a]

[a] *NRC "KI" Petersburg Nuclear Physics Institute, Gatchina, 188300 Russia*

[b] *NRC "Kurchatov institute", Moscow, 123182 Russia*

[c] *JSC "SSC RIAR", Dimitrovgrad, 433510 Russia*

**Abstract**

In order to carry out research in the field of possible existence of a sterile neutrino the laboratory based on SM-3 reactor (Dimitrovgrad, Russia) was created to search for oscillations of reactor antineutrino. The prototype of a multi-section neutrino detector with liquid scintillator volume of 350 l was installed in the middle of 2015. It is a moveable inside the passive shielding detector, which can be set at distance range from 6 to 11 meters from the reactor core. Measurements of antineutrino flux at such small distances from the reactor core are carried out with moveable detector for the first time. The measurements carried out with detector prototype demonstrated a possibility of measuring a reactor antineutrino flux in difficult conditions of cosmic background at Earth surface.

* serebrov@pnpi.spb.ru

## 1. Introduction

At present there is a widely spread discussion about possible existence of a sterile neutrino having much less cross-section of interaction with matter , compared, for instance, with that of a reactor electron antineutrino. It is assumed that owing to reactor antineutrino transitions into a sterile state, the effect of neutrino oscillations at a short reactor distance, as well as deficiency of antineutrino flux at long reactor distance are likely to be observed [1,2]. Moreover, sterile neutrino can be regarded as a candidate for the dark matter.

The idea of neutrino oscillation can be researched by performing the direct measurements of antineutrino flux deviation and neutrino spectrum distortion at small reactor distances. The detector should be moveable and sensitive to neutrino spectrum. The goal of our experiment is to confirm or refute, at a certain accuracy level, the possibility of existing of sterile neutrino. To search for neutrino oscillation in sterile state one has to observe the deviation of reactor antineutrino flux. If such a process exists it can be described by the following oscillation equation:

$$P(\tilde{\nu}_e \to \tilde{\nu}_e) = 1 - \sin^2 2\theta_{14}\ \sin^2\left(1.27\frac{\Delta m_{14}^2[eV^2]L[m]}{E_{\tilde{\nu}}[MeV]}\right) \qquad (1),$$

where $E_{\tilde{\nu}}$ is antineutrino energy, $\Delta m_{14}^2$ and $\sin^2 2\theta_{14}$ the unknown oscillation parameters.

To carry out the experiment it is required to perform measurements of antineutrino flux and spectrum at small reactor distances, e.g. 6-12 meters from almost point-like antineutrino source.

We have examined the opportunity of conducting new experiments at research reactors in Russia. Due to some peculiar characteristics of its construction, reactor SM-3 provides the most favorable conditions for conducting an experiment on search for neutrino oscillations at short distances. Advantages of SM-3 reactor are a compact reactor core ($35\times42\times42$ cm$^3$) with high reactor power being equal to 100 MW, as well as a sufficiently short distance (5 m) from the reactor core to the wall of an experimental hall. Besides, the fact of special significance is that an antineutrino beam can be measured within a sufficiently wide range from 6 to 13 meters [3, 4]. However, SM-3 reactor is at the Earth surface, hence cosmic background is the main difficulty with this experiment. The passive shielding from cosmic background provided by concrete floors of the building is estimated to be about 5-10 meters of water equivalent. This paper devoted to study of background suppression with multi-section detector and we present the results of first measurements of neutrino flux dependence on distance in baseline range 6-11 meters.

## 2. The problem with reactor antineutrino detection in the presence of cosmic background at the Earth surface.

We carried out experiments with two detectors: the first model based on PMT-49B photomultiplier has the undivided vessel with volume 400 l of liquid scintillator and the second model based on PMT-9354 photomultiplier has the divided in 16 sections vessel with 350 l of liquid scintillator.

The detector inner vessel 0.9x0.9x0.5 $m^3$ is filled with liquid scintillator doped with Gadolinium (0.1%). The scintillation type detector is based on IBD (inverse beta decay) reaction $\tilde{\nu}_e + p \rightarrow e^+ + n$. Primary signal - the detector registers positron, which energy depends on antineutrino energy and, also, registers 2 annihilation γ-quanta with energy 511 keV each. Delayed signal - neutrons emerging in reaction are absorbed by Gd to form a cascade of gamma quanta with total energy about 8 MeV. The detector system keeps records of two subsequent signals from positron and neutron - so called correlated events.

Most registered events are form cosmic rays and gamma quanta from the surrounding materials. The detector count rate is approximately $10^2$ $s^{-1}$. Neutrino events rate is $2 \cdot 10^{-3}$ $s^{-1}$ that is 5 orders of magnitude less then total count rate, hence the problem is to distinguish neutrino events from detector data. We applied the method of delayed coincidences with time interval of 100 μs and managed to distinguish correlated events with rate of 1 $s^{-1}$. Most of these events are signals from muons stopped inside the detector and absorbed by nuclei with emission of neutrons. Energy discrimination allows us to lower correlated event rate to $0.1 s^{-1}$ [4,5]. Active shielding decreases background by another order of magnitude. However, signal to background ratio remains at level 20-25%, that significantly affects the precision of measurements of reactor antineutrino flux. That fact inspired us to investigate other methods, except for the active shielding, of background suppression: signal form selection and detector sectioning.

## 3. External and Internal Active Shielding of the Detector

The active shielding of neutrino detector consists of external and internal parts with respect to passive shielding, Fig.1. On the roof of the passive shielding above the detector was installed the external active shielding system («umbrella») made of 12 cm thick scintillator plates $0.5x0.5 m^2$. The total area of «umbrella» is $2x3 m^2$. Taking into consideration the fact that the detector area is $0.9 \times 0.9\ m^2$, such an «umbrella» must capture the main muon flux directed into the neutrino detector area. However, that was not enough and we installed another level of shielding – active shielding around the detector. It also consists of 12 cm thick scintillator plates and PMT-49B photomultipliers. The bottom surface of inner active shielding consists of 5 cm thick scintillator plates and PMT-97. Fig. 1 shows an installation scheme.

Detector passive shielding is made of elements based on steel plates $1 \times 2\ m^2$, 10mm thick. Six 10 mm thick lead sheets are attached to these plates. The inner cabin volume is $2x2x8\ m^3$. From the inside the cabin is covered with plates of borated polyethylene 16 cm thick. The total weight of passive shielding is 60 ton, the volume of borated polyethylene is $10\ m^3$. Inside passive shielding on the rails is situated a moveable platform with the antineutrino detector. The step motor provides platform movements along the rails within the range of 6 to 11 meters from the reactor core.

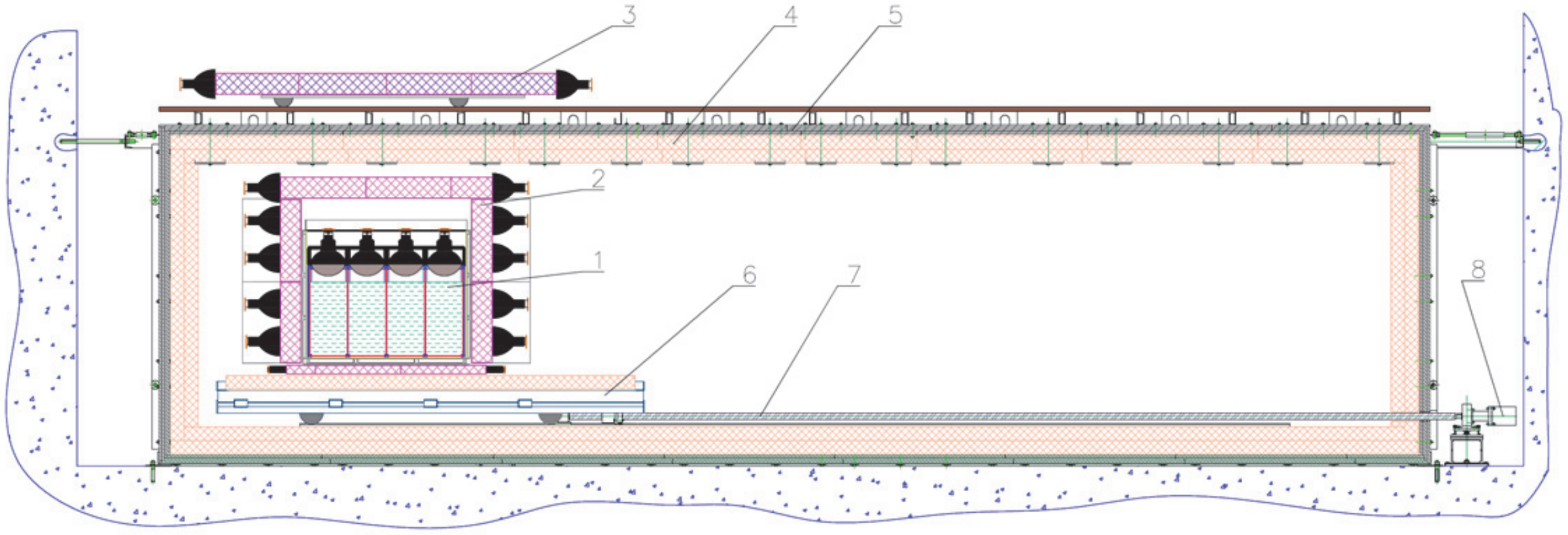

Fig. 1. General scheme of experimental setup. 1 – detector of reactor antineutrino, 2 – internal active shielding, 3 – external active shielding (umbrella), 4 – steel and lead passive shielding, 5 – borated polyethylene passive shielding, 6 – moveable platform, 7 – feed screw, 8 – step motor.

The background suppression (reactor turned off) by active shielding appeared to be by factor of 1.85 averaged by all distances. With external shielding turned on the effect of additional background suppression by internal active shielding appeared to be by factor ~2.0. Hence, the total suppression provided by two level active shielding is by factor ~3.7.

However, signal-background ratio is ~20% (averaged by distance). The signal is the difference between amounts of correlated events in reactor-on and off regimes, and background is amount of correlated events in reactor-off regime. With 20% signal to background ratio the precision of measurements is 3 times less than with zero background conditions. Hence, it is necessary to develop and use additional background suppression methods.

There are at least two ways to solve that problem: make use of special properties of detection either positron or recoil proton. The positron is to annihilate and emit two γ-quanta in opposite directions. For proton detection there is a well-known experimental fact that signals are different for light and heavy particles.

From our research we concluded that signal form selection could be used only if photomultiplier detect the signal from events occurred at small depth, i.e. when radiation responsible for the slow term reach the PMT.

Due to failure of using the method of signal form selection for recoil protons and positrons, we had to consider another option – positron annihilate and emits two γ-quanta in opposite directions. Hence, another option is two detect two γ-quanta with energy 511 keV. In order to fulfill that idea we considered the possibility of creating multi-section detector.

## 4. Multi-section antineutrino detector model

Multi-section model was designed especially for detecting positron emitted in inverse beta decay reaction. As mentioned before, rapid neutrons from cosmic rays are the main problem for Earth-surface experiments. Neutron scattering imitate neutrino reaction. The recoil proton mimics the prompt signal from positron. The delayed signal emits during neutron capturing by Gd in both reactions. The difference in prompt signals is that in neutrino process two γ-quanta are emitted due to positron annihilation. The recoil proton path with high probability lies in single section. 511 keV γ-quanta can be detected in neigbouring sections.

However, with section size 22.5 x 22.5 x 50 $cm^3$, about 70% prompt signals from neutrino events can be detected in single section. Hence, only ~30% of neutrino events are multi-section due to γ-quanta detection in neighbouring sections with respect to section where positron annihilated. The fact that event statistics is 3 times less if we consider only multi-section events is hardly acceptable, so we considered data analysis model with both multi and single section events. Neutrino-like events selection criteria is the ~30% to ~70% ratio for multi and single section events. Hence, if the signal difference for reactor turning on (off) is in ~30% to ~70% ratio for multi and single section events then we consider it to be neutrino signal. This ratio depends on detector inner structure – sections amount and size. Therefore, the precise Monte-Carlo calculations for certain structure are required.

Calculations were performed for detector model described below. The detector consists of 16 sections 22.5 x 22.5 x 50 $cm^3$ with fixed baffles. The baffles are made of acrylic plastic and have specular surface serve to prevent light from leaving the section. Scintillator material is mineral oil ($CH_2$) with addition of Gadolinium 1g per litter. Scintillator light yield is $10^4$photons/MeV. Baffle thickness was not taken into consideration in Monte-Carlo modeling. The air layer between scintillator and PMT increases section energy resolution due to the fact that total internal reflection decrease the solid angle of light yield in PMT close region hence align the photo-collection conditions with respect to altitude. Photons have mirror reflection from the walls. Exponential photon path length in scintillator is 4 meters. Photon reflection probability is 95%. Table 1 represents the probability of double-start events registration when the main signal corresponds to various detector sections. The positron energy considered to be 4 MeV.

Table 1. The probability of double-start events registration in various sections.

| central section | side section | corner section | average |
|---|---|---|---|
| 0.424 | 0.294 | 0.188 | 0.300 |

The probability of double-start registration depends on the section location: central, side or corner. The average probability of double-start event registration for all detector is 30%. Presented probability distribution is a peculiar characteristic of neutrino events, as it is determined by process of positron annihilation with emission of two 511 keV γ-qanta.

Before getting down to the main measurements with a new detector model, measurements with neutron Pu-Be source were made. This experiment proved that rapid neutrons provide single section starts.

Differences in count rates in reactor on and off regimes for double and single starts integrated over all distances are (35±7)% and (65±11)%. With considered precision this ratio allows us to regard registrated events as neutrino-like events. We would like to notice, that multi-section detector makes it possible to carry out additional analysis of events due to registration of relative position of prompt and delayed signals.

Fig. 2 presents the first measurements of dependence of antineutrino flux on the reactor distance (6-11 meters), measurements carried out with sectioned neutrino detector prototype.

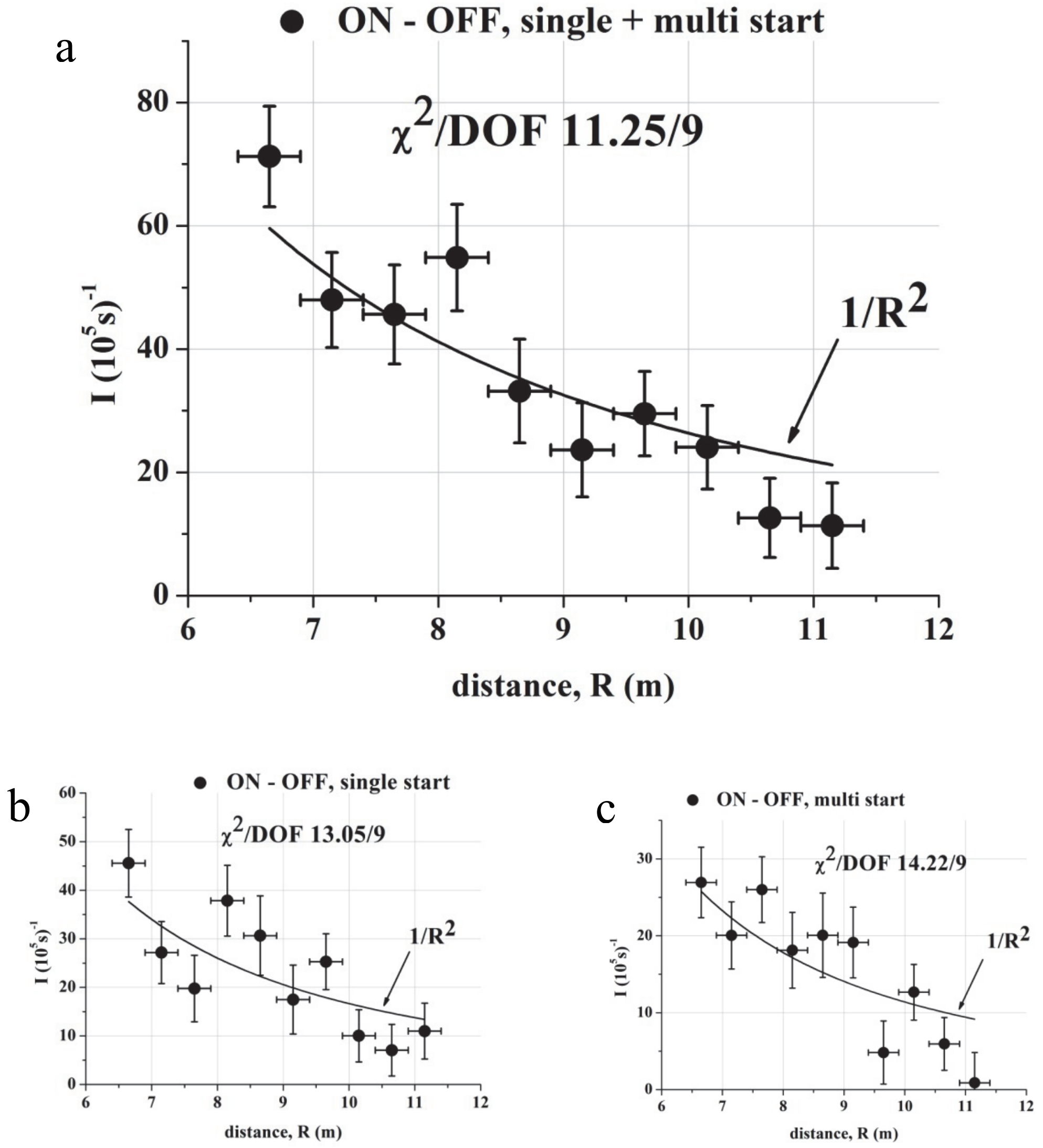

Fig. 2. Neutrino flux dependence on the distance from the reactor core (6-11 meters). Measurements performed with sectioned neutrino detector prototype. Difference in count rates, in reactor on and off regimes, for delayed coincidence in time window 100 μs with cut off random coincidence: a - all events, b - single section prompt signal events, c - multi-section prompt signal events.

The sectioned detector structure allows us to present the distance dependence with 0.5 meter step. The technique of making measurements was to move detector for 1 meter starting with the end position. On the second stage the measurements were repeated with translation of starting position for 0.5 meters. Thus, both halves of the detector measured the same point, averaging in this way somewhat different recording efficiency of each halves of the detector. The measurements with nonsectioned detector were performed with 1 meter steps, hence, the distance dependence consisted of 5 points. In order to increase statistical precision of measurements the results for both detector types are combined. The 5 points dependence is shown in Fig. 3.

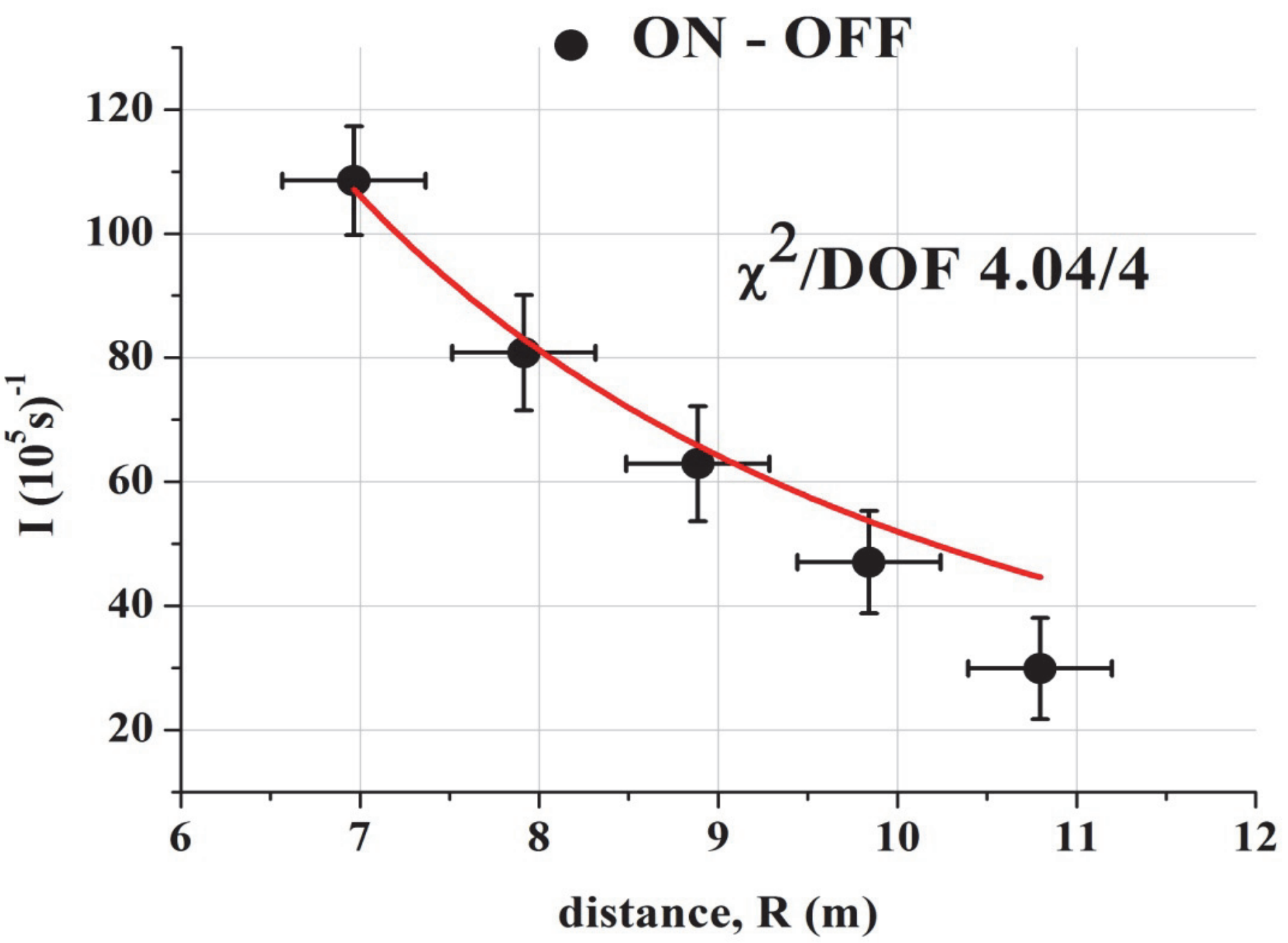

Fig. 3. Measurements of antineutrino flux dependence on distance from the reactor core (6 -11 m) with both prototypes of a neutrino detector.

It should be noted, that measurements of antineutrino flux at such small distances from the reactor core and with moveable detector are carried out for the first time. The results are presented in Fig. 4 as the deviation from $1/R^2$ law, where R – the distance from the reactor core to detector center. The results of all other measurements in close region are also presented in Fig. 4. The results of all other measurements in close reactor zone are taken from [2] and also presented at Fig. 4. To describe the possible effects we present the calculations for four neutrino with various parameters $\Delta m^2_{14}$ and $\sin^2 2\theta_{14}$. This examples lead to conclusion that the measured region is insensitive for oscillation if $\Delta m_{14}^2$>2 eV$^2$. The oscillations occur in unreachable zone in reactor bio protection. If $\Delta m_{14}^2\approx 0.6$ eV$^2$ then there is a broad minimum in measured region (6-11 meters) and the sensitivity of relative measurements significantly decrease. However, in this condition it is necessary to study neutrino spectrum distortion. The highest sensitivity occurs with $\Delta m_{14}^2\approx 1.0$ eV$^2$ and $\Delta m_{14}^2\approx 0.3$ eV$^2$. The required accuracy for each measured point is ~2%.

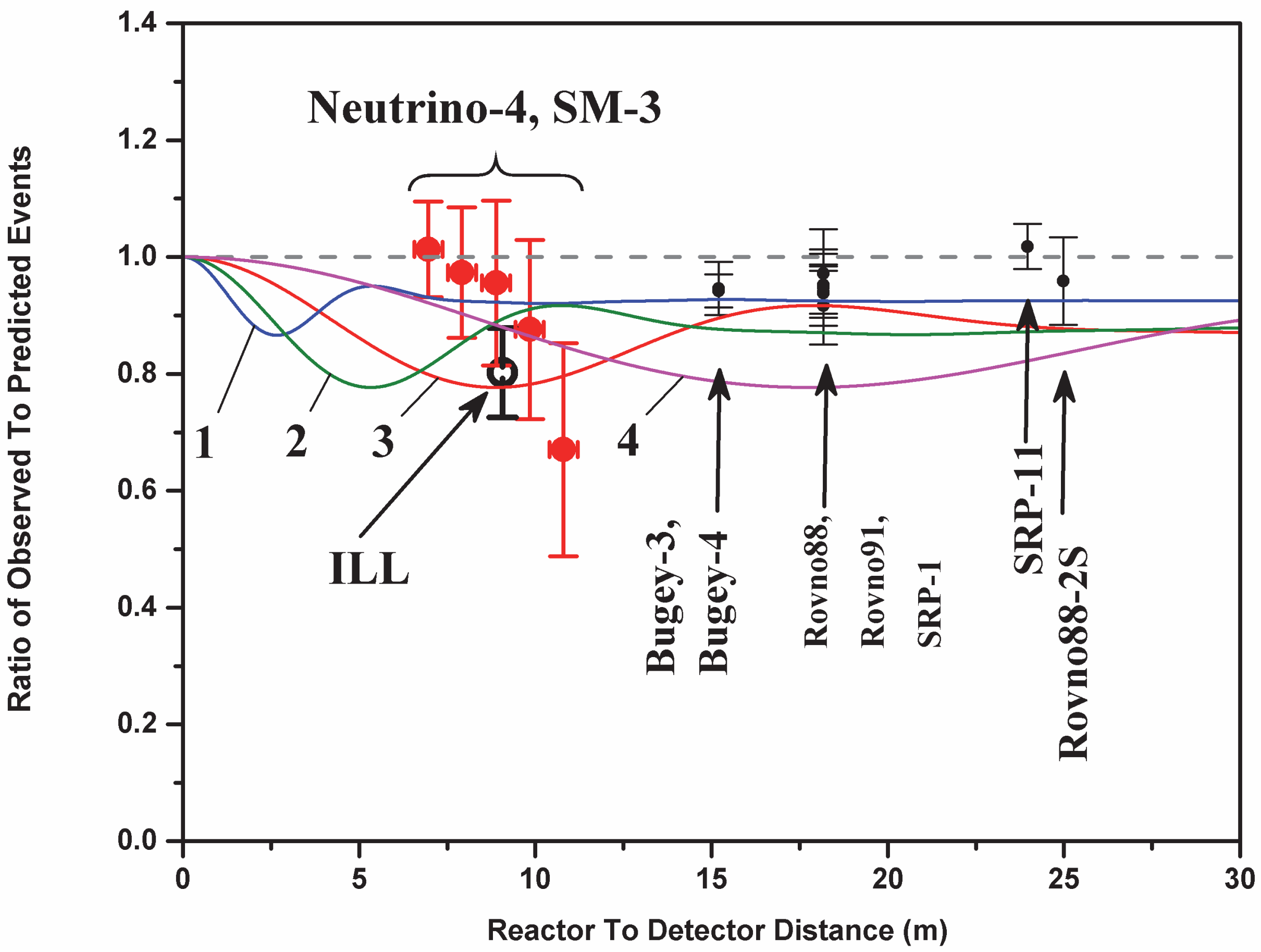

Fig. 4. Measurements of neutrino flux dependence on the distance from the reactor core performed with the prototype of moveable detector in NEUTRINO-4 experiment, and other measurements performed with fixed detectors. 1 – the oscillation curve in scheme with 3 neutrino and 1 sterile neutrino with parameters $\Delta m_{14}^2$=2 eV$^2$, $\sin^2 2\theta_{14}$=0.15; 2 – $\Delta m_{14}^2$=1 eV$^2$, $\sin^2 2\theta_{14}$=0.25; 3 – $\Delta m_{14}^2$=0.6 eV$^2$, $\sin^2 2\theta_{14}$=0.25; 4 – $\Delta m_{14}^2$=0.3eV$^2$, $\sin^2 2\theta_{14}$=0.25.

Among all previous measurements the smallest distance from reactor core was 9 meters in fixed detector experiment under reactor of Institute Laue-Langevin (ILL) [6]. 15 meters distance was used in measurements performed at nuclear power plant in Bougey (France)[7,8]. 18 meters – at reactor of nuclear power plant in Rovno (Ukraine)[9,10], and also at reactor in the Savannah River Site (USA) [11]. 24 meters - the Savannah River Site (USA) [11], 25 meters – Rovno nuclear power plant [9]. All these experiments were performed with fixed detectors designed to measure absolute antineutrino flux and normalized on calculated antineutrino flux from the reactor. The detectors were installed under concrete protection of nuclear plants (about 30 meters of water equivalent [12]). Muon background was suppressed by biological protection of the reactor. For instance, in the Savannah River Site experiment muon flux was suppressed by the factor of 4.3 relatively to muon flux at Earth surface. Under these conditions, the accuracy of detector efficiency with respect to antineutrino registration can possibly reach 5-10%, particularly taking into account that antineutrino flux from nuclear plants is an order of magnitude higher than SM-3 antineutrino flux. Therefore, the signal/background ratio is greater than one in nuclear plant experiments. In our experiment, due to difficulties with carrying out measurements in presence of cosmic rays near earth surface and signal/background ratio about 25-30%, it is hard to reach so high detector efficiency.

The advantage of our experiment is the precision of relative measurements owing to usage of the same detector in search for oscillation effect. Our goal is to find deviation from $1/R^2$ law. Nevertheless, we plan to carry out detailed Monte-Carlo calculations of detector efficiency. According to preliminary estimates calculated efficiency of the multi-section detector has to be 30-35%, and the experimental estimates shows about 20-25%. To consider this divergence as confirmation of reactor antineutrino anomaly it is still premature. In Fig. 4 points of antineutrino flux dependence on distance are put with assumption that both calculated and experimental efficiency of detector registration coincide. We assume that all points presented in Fig. 4 can be moved down if we consider the real detector efficiency. With current statistical accuracy one cannot make any conclusions about neutrino oscillations at small reactor distances.

The required precision of measurements (2%) is not yet achieved, as the researches were carried out only with detector prototype. The full-scale detector with liquid scintillator volume of 3 $m^3$ (5x10 sections) is at preparation stage. New detector will allow us to obtain up to 1.5-3.0% statistic accuracy of measurements at the distances from 6 to 12 meters after 2 years of measurements, and we hope that would bring light to the problem of neutrino oscillation in the sterile state.

As it was noticed in introduction, besides observing of the spatial flux variations it is important to study neutrino spectrum in various detector positions. This task requires detector energy calibration. Before the experiment have started the multi-section model had been calibrated with use of γ-quanta source ($^{22}Na$) and neutron source (Pu-Be). The results of energy calibration are presented in Fig. 5.

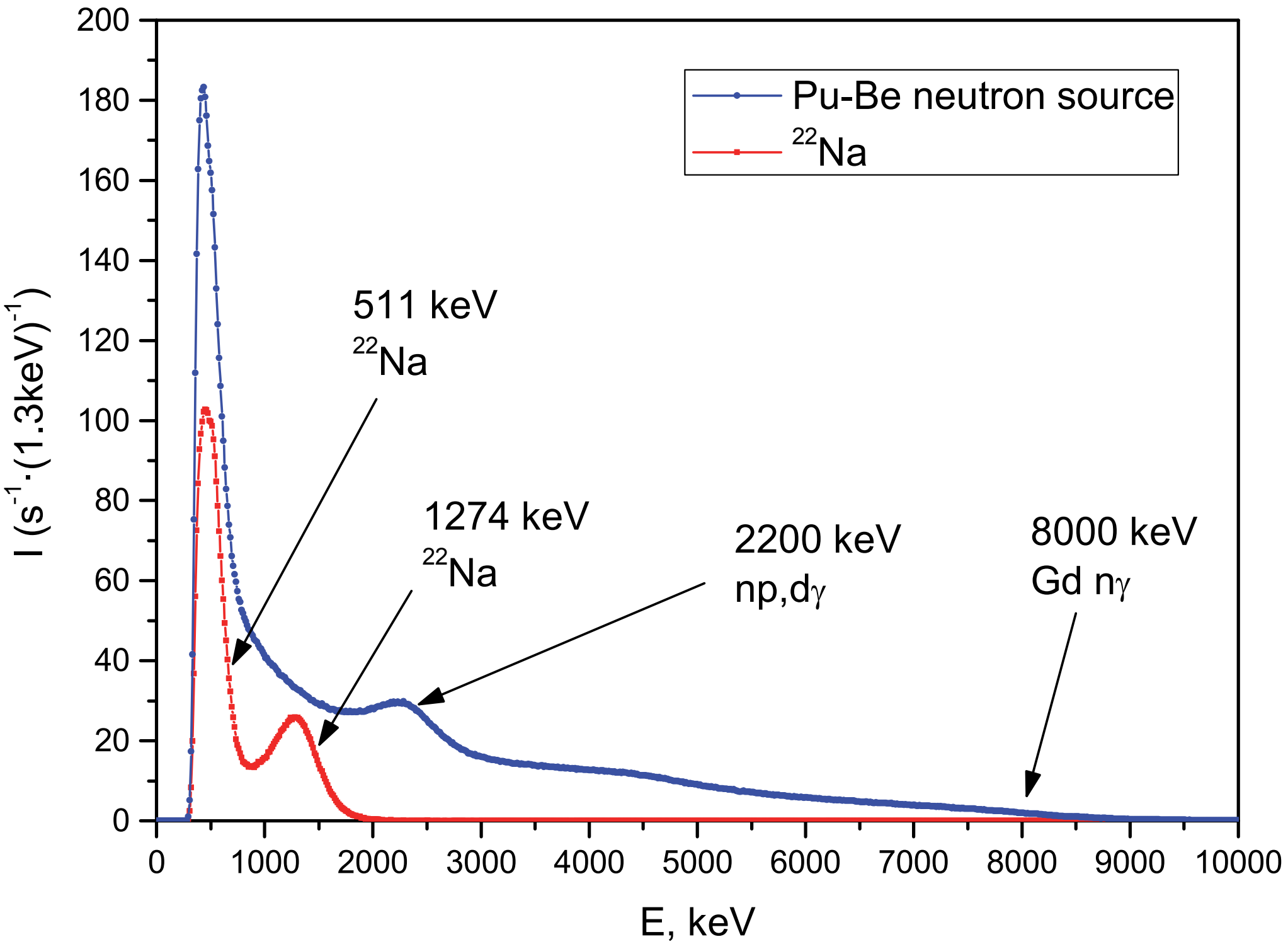

Fig. 5. Results of energy calibration.

The difference spectra (ON-OFF) of prompt signals with 5 distance points are presented in Fig. 6 in compare with Monte-Carlo calculations. The measured integral and calculated one is equal each other to observe variation of spectrum form.

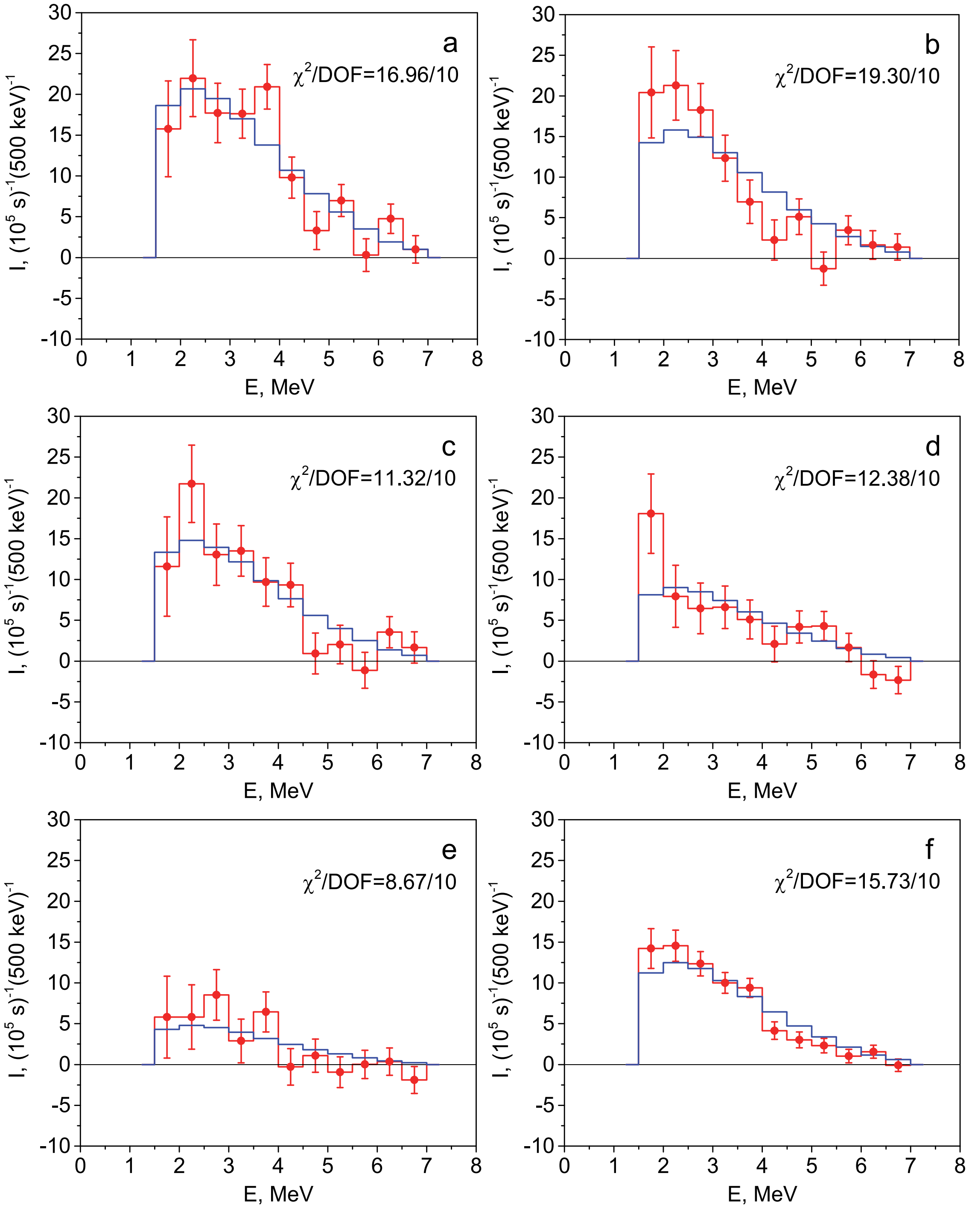

Fig. 6. Results of spectrum measuring at various distances and comparison with MC calculations of zero oscillation scenario: a – 6.9 m, b – 7.9 m, c – 8.9 m, d – 9.9 m, e – 10.9 m, f –averaged spectrum.

To observe spectrum distortion even more statistics are required than to observe spatial flux deviations. The results obtained with detector prototype are presented here only for illustration. It is possible to extrapolate these results to conclusion that at full-scale experiment statistical uncertainties will be less by factor 5-7.

## Conclusion

Measurements of antineutrino flux at small reactor distances with moveable detector carried out for the first time. The measurements conducted with multi-section detector prototype pointed out that in difficult conditions of cosmic background at Earth surface it is still possible to carry out measurements of antineutrino flux, but it requires a large detector to obtain necessary precision.

The authors are grateful to the Russian Foundation for Basic Research for support under Contract No. 14-22-03055-ofi_m. The delivery of the scintillator from the laboratory leaded by Prof. Jun Cao (Institute of High Energy Physics, Beijing, China) has made a considerable contribution to this research.